# Zitterbewegung by Quantum Field Theory Considerations

Zhi-Yong Wang, Cai-Dong Xiong

*School of Optoelectronic Information, University of Electronic Science and Technology of China,*

*Chengdu 610054, CHINA*

The validity of the work by Lamata *et al* [Phys. Rev. Lett. **98**, 253005 (2007)] can be further shown by quantum field theory considerations.

In a recent interesting Letter [1], Lamata *et al* have proposed the simulation of the Dirac equation for a free spin-1/2 particle in a single trapped ion, by which they have studied relevant quantum-relativistic effects such as the zitterbewegung (ZB). However, it is still vague whether the ZB phenomenon also exists at the level of quantum field theory. For example, in Ref. [1], there is a statement that "its (i.e., the ZB effect's) existence is even questioned by quantum field theory considerations" (in the top left hand side of page 3). Moreover, in Ref. [2], the Abstract contains a statement that "we also find that quantum field theory prohibits the occurrence of ZB for an electron", however, this statement is very misleading, because what the authors genuinely means is that "a quantum field theory that *neglects the electron-positron interactions* prohibits any ZB", which is equivalent to the conclusion that a quantum mechanics that *neglects the interference between the positive- and negative-energy components of wavefunctions* prohibits any ZB, and then it is just a well-known and trivial conclusion. As a result, from their work one cannot know whether quantum field theory genuinely prohibits the occurrence of ZB.

In fact, seeing that a hole can be interpreted as a positron, the physical argument for the ZB of an electron (in a bound or free state) within the framework of the hole theory [3, 4] can be restated in terms of quantum field theory as follows: around an original electron, virtual electron-positron pairs are continuously created (and annihilated subsequently) in the vacuum, and there exists an exchange interaction and a related scattering, i.e., the original electron can annihilate with the positron of a virtual pair, while the electron of the virtual pair which is left over now replaces the original electron, by such an exchange interaction the ZB occurs. That is, quantum field theory permits the occurrence of the ZB for an electron, and it shows that the ZB arises from the influence of virtual electron-positron pairs (or vacuum fluctuations) on the electron. On the other hand, relativistic quantum mechanics is only a transitional theory to quantum field theory, then for completeness we will further develop the work in Ref. [1] at the level of quantum field theory, which is worthy of being done because the interest in ZB is recently rekindled by the investigations on spintronic, graphene, and superconducting systems, etc. [5-9]. In particular, there are recently important progresses in improving the predictions for detecting ZB and relating them to Schrodinger cats in trapped ions [10, 11].

According to Ref. [1], applying four metastable ionic internal states $|a\rangle$, $|b\rangle$, $|c\rangle$ and $|d\rangle$, the general state can be expressed as $|\psi\rangle = \psi_a|a\rangle + \psi_b|b\rangle + \psi_c|c\rangle + \psi_d|d\rangle$, where the coefficients $\psi_i$'s ($i = a,b,c,d$) represent the probability amplitudes lying in the ionic internal level $|i\rangle$, they satisfy $\sum_i |\psi_i|^2 = 1$ provided that there are only the four internal states. The quantity $\langle j|i\rangle$ ($i, j = a,b,c,d$, $i \neq j$) represents the transition probability amplitudes from $|i\rangle$ to $|j\rangle$, and the matrix representation of the state vector $|\psi\rangle$ is the four-component Dirac bispinor. Further, in the field-quantized sense, let the *vacuum state*

$|0\rangle$ represent the state where the four ionic internal levels are unoccupied, and define creation operators $a^\dagger$, $b^\dagger$, $c^\dagger$ and $d^\dagger$ by $i^\dagger|0\rangle = |i\rangle$ ($i = a,b,c,d$), then the field operator associated with the four ionic internal levels can be defined as

$$\psi := a^\dagger \psi_a + b^\dagger \psi_b + c^\dagger \psi_c + d^\dagger \psi_d. \tag{1}$$

Obviously, one has $|\psi\rangle = \psi|0\rangle$. Likewise, the Schrödinger equation of the field operator can be expressed as $i\partial \psi(t,\bm{x})/\partial t = (\bm{\alpha}\cdot\hat{\bm{p}} + \beta m)\psi(t,\bm{x})$ (the natural unit $\hbar = c = 1$ is applied, according to Ref. [1], $c \equiv 2\eta\Delta\tilde{\Omega}$, $m \equiv \hbar\Omega/c^2$, position and momentum are associated with the respective ionic variables), it is isomorphic equivalent to the free Dirac equation, such that as the general solution of the Schrödinger equation, the field operator $\psi$ can be rewritten as (in contrast to Ref. [1], here the position vector is $\bm{x}$ instead of $\bm{r}$):

$$\psi(t,\bm{x}) = \sum_{\bm{p},s} \sqrt{m/VE}[e(\bm{p},s)u(\bm{p},s)\exp(-ip\cdot x) + f^\dagger(\bm{p},s)v(\bm{p},s)\exp(ip\cdot x)], \tag{2}$$

where $\int d^3\bm{x} = V \to +\infty$, $p.x = Et - \bm{p}\cdot\bm{x}$, $E = \sqrt{\bm{p}^2 + m^2}$, $f^\dagger$ denotes the Hermitian adjoint of $f$, $u$ and $v$ are the Dirac spinors in the momentum space:

$$u(\bm{p},s) = \sqrt{\frac{E+m}{2m}}\begin{pmatrix} \chi_s \\ \dfrac{\bm{\sigma}\cdot\bm{p}}{E+m}\chi_s \end{pmatrix},\quad v(\bm{p},s) = \sqrt{\frac{E+m}{2m}}\begin{pmatrix} \dfrac{\bm{\sigma}\cdot\bm{p}}{E+m}\chi_s \\ \chi_s \end{pmatrix}, \tag{3}$$

where $\chi_s$'s satisfy the orthonormality and completeness relations: $\chi_s^\dagger \chi_{s'} = \delta_{ss'}$, $\sum_s \chi_s \chi_s^\dagger = I_{2\times2}$ ($I_{2\times2}$ is the 2×2 unit matrix), $\bm{\sigma} = (\sigma_x, \sigma_y, \sigma_z)$ is the Pauli matrix-vector. In contrast to the spin-1/2 particles in Dirac theory, we call $s = \pm 1/2$ the pseudospin indices, and call $e^\dagger$ and $e$ ($f^\dagger$ and $f$) the creation and annihilation operators of quasiparticles (anti-quasiparticles), respectively. Let $|n_l\rangle$ ($l = e, f$) be the state containing $n_l$ quasiparticles (or anti-quasiparticles), one has $l^\dagger l |n_l\rangle = n_l |n_l\rangle$, and as fermi excitations, one has $n_l = 0$ or 1. Eqs. (1) and (2) imply that $e$ and $f^\dagger$ are the linear combinations of

$i^\dagger$ ($i = a, b, c, d$) and vice versa, these quasiparticles and anti-quasiparticles can be regarded as elementary excitations corresponding to a combination of the four ionic internal levels.

In quantum field theory, the usual position operator is no longer a well-defined concept, but one can study ZB via the current vector $\boldsymbol{I} = \int \psi^\dagger \boldsymbol{\alpha} \psi \mathrm{d}^3 \boldsymbol{x}$. One can prove that $\boldsymbol{I} = \boldsymbol{v} + \boldsymbol{z}_\perp + \boldsymbol{z}_\parallel$, where $\boldsymbol{v} = \sum_{p,s}(\boldsymbol{p}/E)[e^\dagger(\boldsymbol{p},s)e(\boldsymbol{p},s) - f^\dagger(\boldsymbol{p},s)f(\boldsymbol{p},s)]$ is the classical term, while $\boldsymbol{z}_\perp$ and $\boldsymbol{z}_\parallel$ are the transverse and longitudinal ZB currents (respectively perpendicular and parallel to the momentum vector $\boldsymbol{p}$):

$$\boldsymbol{z}_\perp = \sum_{\boldsymbol{p}} \{\sqrt{2}\boldsymbol{\eta}_+[e^\dagger(\boldsymbol{p},-1/2)f^\dagger(-\boldsymbol{p},1/2)\exp(\mathrm{i}2Et) + h.c.] + h.c.\}, \quad (4)$$

$$\boldsymbol{z}_\parallel = \sum_{\boldsymbol{p}} (m/E)\boldsymbol{\eta}_\parallel \{[e^\dagger(\boldsymbol{p},1/2)f^\dagger(-\boldsymbol{p},1/2) - e^\dagger(\boldsymbol{p},-1/2)f^\dagger(-\boldsymbol{p},-1/2)]\exp(\mathrm{i}2Et) + h.c.\}, \quad (5)$$

where $h.c.$ denotes the Hermitian conjugate of the preceding term, to obtain Eqs. (4) and (5) the relation $\sum_{\boldsymbol{p}=-\infty}^{\boldsymbol{p}=+\infty} F(\boldsymbol{p}) = \sum_{\boldsymbol{p}=-\infty}^{\boldsymbol{p}=+\infty} F(-\boldsymbol{p})$ is applied, $\boldsymbol{\eta}_\pm$ and $\boldsymbol{\eta}_\parallel$ represent the three orthonormal eigenvectors of the spin-projection operator of vector fields, i.e., $\boldsymbol{\eta}_\parallel$ is the longitudinal polarization vector of the spin-1 field, while $\boldsymbol{\eta}_+$ and $\boldsymbol{\eta}_-$ are the right- and left-hand circular polarization vectors, respectively, they are

$$\begin{cases} \boldsymbol{\eta}_+ = (1/\sqrt{2}|\boldsymbol{p}|)(\dfrac{p_1 p_3 - \mathrm{i}p_2|\boldsymbol{p}|}{p_1 - \mathrm{i}p_2}, \dfrac{p_2 p_3 + \mathrm{i}p_1|\boldsymbol{p}|}{p_1 - \mathrm{i}p_2}, -(p_1 + \mathrm{i}p_2)) \\ \boldsymbol{\eta}_- = (1/\sqrt{2}|\boldsymbol{p}|)(\dfrac{p_1 p_3 + \mathrm{i}p_2|\boldsymbol{p}|}{p_1 + \mathrm{i}p_2}, \dfrac{p_2 p_3 - \mathrm{i}p_1|\boldsymbol{p}|}{p_1 + \mathrm{i}p_2}, -(p_1 - \mathrm{i}p_2)) \\ \boldsymbol{\eta}_\parallel = \dfrac{\boldsymbol{p}}{|\boldsymbol{p}|} = \dfrac{1}{|\boldsymbol{p}|}(p_1, p_2, p_3) \end{cases} \quad (6)$$

Let $\boldsymbol{p} = (0, 0, p_z)$ with $p_z \geq 0$, one has $\boldsymbol{\eta}_\pm = (1/\sqrt{2})(1, \pm\mathrm{i}, 0)$ and $\boldsymbol{\eta}_\parallel = (0, 0, 1)$.

According to the interpretation of the Dirac ZB and our isomorphic equivalent relations, the ZB here can be attributed to the influence of continuously virtual transition processes between the four ionic internal levels, i.e., the influence of continuously creating and annihilating virtual pairs of elementary excitations. In fact, Eqs. (4) and (5) show that the ZB currents are related to the creation and annihilation operators of elementary excitation pairs, i.e., $e^\dagger f^\dagger$ and $ef$.

In a word, the work in Ref. [1] is also valid by quantum field theory considerations, and at the level of quantum field theory, the ZB is due to the influence of continuously *virtual* transition processes between the four ionic internal levels.

**Acknowledgments**: The first author (Z. Y. Wang) would like to thank professor Erasmo Recami for his useful discussions. This work was supported by the National Natural Science Foundation of China (Grant No. 60671030) and by the Scientific Research Starting Foundation for Outstanding Graduate, UESTC, China (Grant No. Y02002010501022).